# Zero-field spin waves in YIG nano-waveguides

K. O. Nikolaev[1,‡], S. R. Lake[2,‡], G. Schmidt[2,3], S. O. Demokritov[1], and V. E. Demidov[1]

[1]*Institute of Applied Physics, University of Muenster, 48149 Muenster, Germany*

[2]*Institut für Physik, Martin-Luther-Universität Halle-Wittenberg, 06120 Halle, Germany*

[3]*Interdisziplinäres Zentrum für Materialwissenschaften, Martin-Luther-Universität Halle-Wittenberg, 06120 Halle, Germany*

Spin-wave based transmission and processing of information is a promising emerging nano-technology that can help overcome limitations of traditional electronics based on the transfer of electrical charge. Among the most important challenges for this technology is the implementation of spin-wave devices that can operate without the need for an external bias magnetic field. Here we experimentally demonstrate that this can be achieved using sub-micrometer wide spin-wave waveguides fabricated from ultrathin films of low-loss magnetic insulator – Yttrium Iron Garnet (YIG). We show that these waveguides exhibit a highly stable single-domain static magnetic configuration at zero field and support long-range propagation of spin waves with gigahertz frequencies. The experimental results are supported by micromagnetic simulations, which additionally provide information for optimization of zero-field guiding structures. Our findings create the basis for the development of energy-efficient zero-field spin-wave devices and circuits.

‡ : authors that contributed equally to the work

Waves of magnetization (spin waves) possess wavelengths from micrometers to nanometers in the gigahertz frequency range, and can be guided in thin-film magnetic nanostructures. This makes them an ideal candidate for implementation of high-speed, wave-based computing and signal processing at the nanoscale, with the addendum that they potentially may outperform traditional semiconductor technology in terms of footprint and energy efficiency[1-3]. The latter advantage has become particularly prominent with the recent developments in fabrication of low-loss nanometer-thick films of magnetic insulator, Yttrium Iron Garnet (YIG)[4-6], which can be structured at the nanometer scale[7-9] and allows long-distance propagation of spin waves with characteristic decay lengths of tens of micrometers[10-12] significantly exceeding those in metallic ferromagnets[13]. As a result, transmission of signals using spin waves in YIG nanostructures can be more energy efficient compared to transmission based on electrical charge transfer[1,2].

Although spin-wave circuits are a promising alternative to traditional semiconductor electronic circuits, there are still significant challenges that need to be addressed to improve their practical attractiveness. One of the most significant challenges is to eliminate the requirement of static bias magnetic field to operate spin-wave devices. Therefore, the community has invested significant efforts toward developing the capability to transmit spin waves at zero field[14-22]. In particular, it was shown that zero-field transmission can be achieved using domain walls and spin textures as propagation channels for spin waves[14-19]. Another way to eliminate the static field requirement is to use materials and structures exhibiting different magnetic anisotropies[20-22]. However, while zero-field propagation has been experimentally confirmed in these works, the decay length of spin waves was found to be relatively short due to the large magnetic losses in the materials used. Perpendicular magnetic anisotropy, which enables zero-field or low-field spin-wave transmission, also can be induced in low-loss YIG films by adding different dopants[23-24]. However, in YIG films with sub-100 nm thickness, such doping inevitably results in a significant increase of losses, diminishing the main advantage of this material.

In this Letter, we study experimentally zero-field propagation of spin waves in nanometer-thick YIG waveguides with sub-micrometer width. We show that the shape anisotropy in such waveguides is strong enough to efficiently stabilize a single-domain static magnetic configuration at zero bias field. Due to the small width of the waveguide, the spectrum of spin waves is located at frequencies above 1 GHz under these conditions. We image the propagating spin waves with spatial and phase resolution, which allows us to confirm that the propagation is not disturbed by any inhomogeneities in the static magnetic configuration. These measurements also allow us to directly determine the wavelength and the decay length of spin waves. The latter is found to be about 20 µm for a 500-nm wide waveguide, which is an excellent value for sub-micrometer YIG conduits[9,12]. We also perform micromagnetic simulations, which show good agreement with the experimental results and allow us to address the effects of geometrical parameters on the propagation characteristics and stability of the single-domain state. Our findings open new perspectives for the implementation of spin-wave devices and circuits that do not require space-consuming elements necessary to create a bias magnetic field, which significantly improves the attractiveness of spin-wave technology for applications.

Figure 1(a) shows the schematics of the experiment. We study the propagation of spin waves in a 500-nm wide and 80-nm thick waveguide patterned by electron-beam lithography and lift-off from a YIG film grown by pulsed laser deposition on a gadolinium gallium garnet (GGG) substrate with <111> orientation[6]. The film has a saturation magnetization $4\pi M_s =$ 1.75 kG and the Gilbert damping parameter $\alpha = 4\times10^{-4}$. A static magnetic field $H$ is applied parallel to the waveguide axis to create a uniform static magnetization configuration and then reduced to 0. The spin waves are inductively excited using a 500-nm wide and 200-nm thick Au stripe antenna carrying microwave electric current with a frequency $f$. The propagation of spin waves is imaged by space- and phase-resolved micro-focus Brillouin light scattering (BLS) spectroscopy[13]. The probing laser light with a wavelength of 473 nm is focused on the YIG surface (Fig. 1a) by a microscope objective lens with NA = 0.9. The light scattered from the spin waves is collected by the same lens and interferes with the initial light modulated by the microwave signal used to excite spin waves. This allows us to obtain information about the phase of the waves at the point of observation. The light is then analyzed using a six-pass Fabry-Perot interferometer. The resulting signal is proportional to the local amplitude of the spin wave $A\cos(\Delta\varphi)$, where $\Delta\varphi$ is the phase difference between the excitation signal and the wave at a given spatial position. By moving the probing-light spot over the surface of the waveguide, we record spatial maps of the spin-wave amplitude.

A representative two-dimensional map measured at $H = 0$ and $f = 1.5$ GHz by moving the laser spot over an 0.5×19 µm$^2$ area is shown in Fig. 1b. The shown data clearly indicate that high-frequency spin waves can propagate in the studied waveguide even in the absence of a bias magnetic field. Additionally, the map shows no signatures of any local phase distortions, indicating that the static magnetization in the waveguide remains uniform after the field is removed. Figure 1c shows a section of the map along the axis of the waveguide ($z = 0$ corresponds to the middle of the antenna). We fit the experimental data (symbols) by the $ae^{-x/\kappa}\cos(kx + b)$ function (solid curve) to determine the wavenumber $k$ and the decay length $\kappa$ of the spin wave at a given frequency $f$. The results obtained for different frequencies are summarized in Fig. 2, where the symbols show the experimental data and the curves show the data obtained from micromagnetic simulations (see below). The dispersion spectrum (Fig. 2a) shows that at $H = 0$, the frequencies of spin waves lie in the range 1.35-1.75 GHz. The dispersion curve starts at $k = 0$ at $f = 1.75$ GHz. As the wavenumber increases, the frequency decreases, as expected for the so-called backward volume spin waves propagating parallel to the static magnetization[25]. Note that the lowest frequency in the experimentally observed spin-wave band is determined by the shortest wavelength of 0.8 µm ($k \approx 8$ µm$^{-1}$), which can be excited by the antenna used. Over the significant part of the spin-wave frequency band, the decay length is equal to about 20 µm (Fig. 2b). It starts to decrease close to the boundaries of the band due to a decrease in the group velocity, which is determined by the slope of the dispersion curve (Fig. 2a).

Let us now turn to the question of the stability of the single-domain state of the waveguide with respect to a change in the static magnetic field. We address this question by analyzing spin-wave propagation characteristics at various fields. We vary the static magnetic field in the range $H = \pm 200$ Oe with a step size of 10 Oe and record for each field the intensity of spin waves as a function of the excitation frequency at a distance $z = 10$ µm from the antenna. As a result, we obtain data about the frequencies of the spin-wave band for various $H$. Figures 3a and 3b show these data in a form of a grayscale map in field-frequency coordinates. Figure 3a corresponds to $H$ varying from -200 to 200 Oe, and Fig. 3b corresponds to $H$ varying in the opposite direction. The dark color in the map clearly indicates the position of the spin-wave band in frequency space. When the waveguide is magnetized by $H = -200$ Oe (Fig. 3a), the band is located between 2.15 to 2.5 GHz. As the magnitude of $H$ decreases to 0, the band monotonically shifts to 1.35-1.75 GHz (see also Fig. 2a) and continues to shift down even when $H$ changes its direction ($H = 0$-80 Oe). Note that, in this range, the orientation of the static field is opposite to the orientation of the magnetization in the waveguide. Finally, at $H = 90$ Oe (marked by a vertical dashed line), the band abruptly shifts to higher frequencies, which is likely associated with the switching of the static magnetization in the waveguide to a state parallel to the magnetic field. The variation of the field in the opposite direction (Fig. 3b) shows very similar behaviors with the switching occurring at $H = -90$ Oe. We emphasize that the phase-resolved measurements over the entire range $H = \pm 200$ Oe reveal smooth propagation of spin waves without any phase perturbations, which indicates that the waveguide remains in the single-domain state at all fields.

To support our interpretation, we perform micromagnetic simulations using the package mumax3[26]. We consider a waveguide with a thickness of 80 nm and lateral dimensions of 0.5 by 20 μm. The computational domain is split into 10×10×10 nm³ cells. The exchange constant is set to a standard for YIG value of 3.66 erg/cm. The saturation magnetization and the damping parameter are set according to the known values: $4\pi M_s$ = 1.75 kG, $\alpha = 4\times 10^{-4}$.

First we perform simulations of the static magnetic state. Following the procedure used in the experiment, we magnetize the waveguide by applying a static field $H$ = -200 Oe and then vary $H$ through zero to 200 Oe in 5 Oe increments, allowing the system to relax to a stable state for each $H$ (in these calculations, we use $\alpha = 1$ to speed up the convergence). In agreement with the experimental data, at all fields, we observe a single-domain state with the magnetization oriented along the waveguide axis, except for small regions at the ends of the waveguide. Figure 3c shows the normalized dependence of the z-component of the static magnetization averaged over the volume of the waveguide vs $H$ – the hysteresis curve. As seen from these data, the hysteresis curve is perfectly rectangular, which indicates the absence of multi-domain states. The found coercive field $H_c$ = 100 Oe is also in a good agreement with the experimental data (Figs. 3a and 3b).

Next, we simulate the propagation of spin waves. We excite waves by applying a single-frequency out-of-plane dynamic magnetic field at the center of the waveguide and determine the wavelength and the decay length of the excited waves. The dispersion curves obtained from calculations at different $H$ are shown in Fig. 3d (solid curves) together with the experimental data (symbols). As seen from these data, the simulations reproduce the experimental results well (see also the data for propagation length in Fig. 2b). Note that good agreement is also achieved for the state, when the static magnetic field is anti-parallel to the magnetization in the waveguide (data for $H$ = 50 Oe in Fig. 3d).

We now use micromagnetic simulations to analyze the effects of waveguide geometry on the zero-field propagation of spin waves. Figure 4a shows the calculated dispersion curves obtained at $H = 0$ for waveguides with different widths $w$. As seen from these data, with the decrease in $w$, the frequencies of spin waves shift upwards, while the slope of the dispersion curves decreases. To characterize these effects quantitatively, we plot in Fig. 4b the frequency of spin waves at $k = 0$, $f_0$, and the spectral width of the accessible spin-wave band, $\Delta f$, measured as the difference between $f_0$ and the frequency corresponding to a wavelength of 1 µm (marked by a vertical dashed line in Fig. 4a), which approximately corresponds to the smallest wavelength that can be efficiently excited by an inductive antenna. As seen from Fig. 4b, by reducing the width of the waveguide, one can increase the frequency of spin waves in the zero-field regime. This frequency control is most efficient at $w > 150$ nm and strongly decreases at smaller widths. The increase in the frequency $f_0$ with decreasing $w$ is associated with an increase in the effective transverse wavenumber of the spin wave, which is determined by the quantization over the waveguide width[27]. In the small-width regime, the quantization conditions are known to change due to the increasing role of the exchange interaction[28], which leads to saturation of the $f_0(w)$ dependence at $w < 150$ nm. We also note, that in the small-width regime, $\Delta f$ drastically decreases. This not only limits the available bandwidth of spin-wave devices, but also results in small group velocities, leading to short decay lengths of spin waves. These data indicate that the choice of waveguide width is crucial for optimization of zero-field spin-wave devices and should be based on the optimal combination of center frequency and bandwidth required for a particular application.

Finally, we analyze the influence of the waveguide width on the stability of the zero-field state. We perform micromagnetic simulations of hysteresis curves for different $w$ that yield the dependence of the coercive field, $H_c$, on the waveguide width (Fig. 4c). As seen from these data, a reduction of the width results in an increase of the stability of the zero-field state. We associate this with an increase of the field, which is necessary to create domain walls that initiate the magnetization reversal process. This is mainly due to the increasing energy contribution of edge stray fields of magnetic domain walls in narrow waveguides. As a result, for a waveguide with a width of 50 nm, the coercive field reaches 360 Oe. More interestingly, calculations show that in YIG waveguides with a relatively large width of 1000 nm, the coercive field remains as large as 50 Oe. This allows one to use the entire range $w$ = 50-1000 nm for zero-field spin wave devices.

In conclusion, we have shown that YIG-based spin-wave nano-devices do not necessarily require a bias magnetic field for operation. Zero-field operation can be easily achieved by exploiting shape anisotropy in YIG nano-waveguides over a wide range of geometric parameters. Due to spin-wave quantization in waveguides with a sub-micrometer width, typical operating frequencies can be as high as 1-2 GHz even in the absence of a bias field. Additionally, the propagation of spin waves in this regime is characterized by long decay lengths of several tens of micrometers. These observations open new possibilities for implementation of scalable, energy-efficient spin-wave nano-devices and circuits.

This work was supported by the Deutsche Forschungsgemeinschaft (DFG, German Research Foundation) – Project-ID 433682494 – SFB 1459, and TRR227, project B02, WP3.

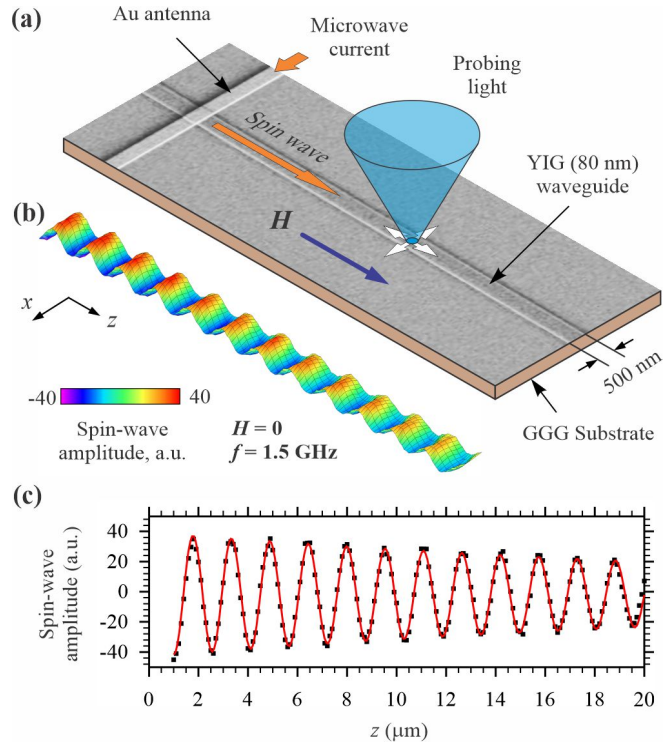

**Figure 1.** Experimental devices and measurement technique. **(a)** Schematics of the experiment. Spin waves in a 500-nm wide and 80-nm thick YIG waveguide are excited using a Au stripe antenna carrying microwave electric current. The waveguide is first magnetized by a static magnetic field $H$, which is then reduced to 0. The propagation of spin waves is imaged by space- and phase-resolved micro-focus BLS spectroscopy. (b) Representative two-dimensional map of the spin-wave amplitude measured at $H = 0$ and $f = 1.5$ GHz. (c) Section of the map along the axis of the waveguide ($z = 0$ corresponds to the middle of the antenna). Symbols show the experimental data. Curve shows a fit by a damped sinusoidal function.

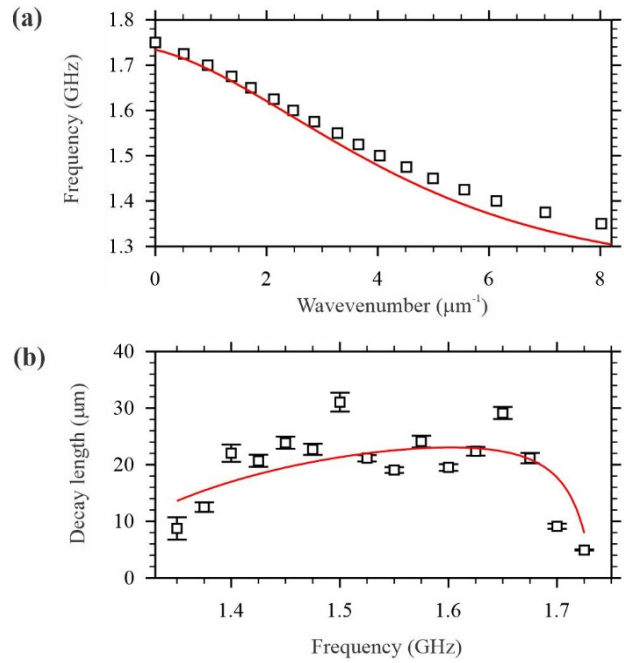

**Figure 2.** Propagation characteristics of spin waves at $H = 0$. (**a**) Dispersion curve. (**b**) Frequency dependence of the decay length. Symbols show the experimental data. Curves show the results of micromagnetic simulations.

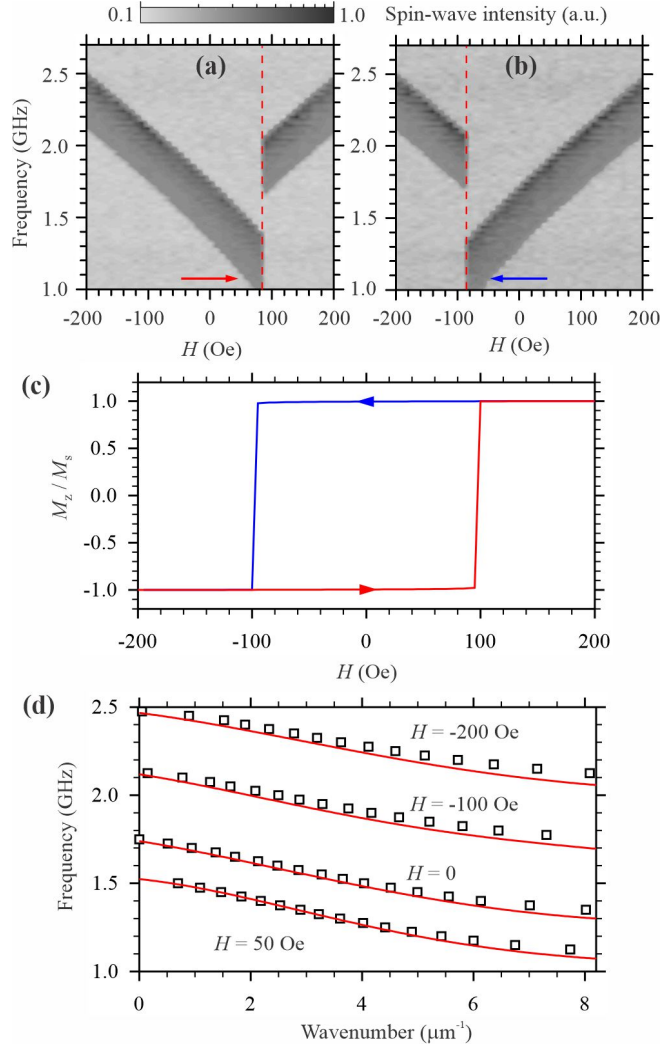

**Figure 3.** Dependence of the propagation characteristics of spin waves and the static magnetic state of the waveguide on the bias magnetic field. (a) and (b) Intensity of spin waves detected at a distance $z = 10$ μm from the antenna in field-frequency coordinates. (a) corresponds to $H$ varying from -200 to 200 Oe, and (b) corresponds to $H$ varying in the opposite direction, as shown by arrows. Vertical dashed lines mark the field, at which the direction of the static magnetization is reversed. (c) Normalized field dependence of the $z$-component of the static magnetization averaged over the volume of the waveguide obtained from micromagnetic simulations. Arrows show the direction in which $H$ changes. (d) Comparison of the dispersion curves of spin waves obtained from the experiment (symbols) with those obtained from micromagnetic simulations (curves) at different $H$, as labelled.

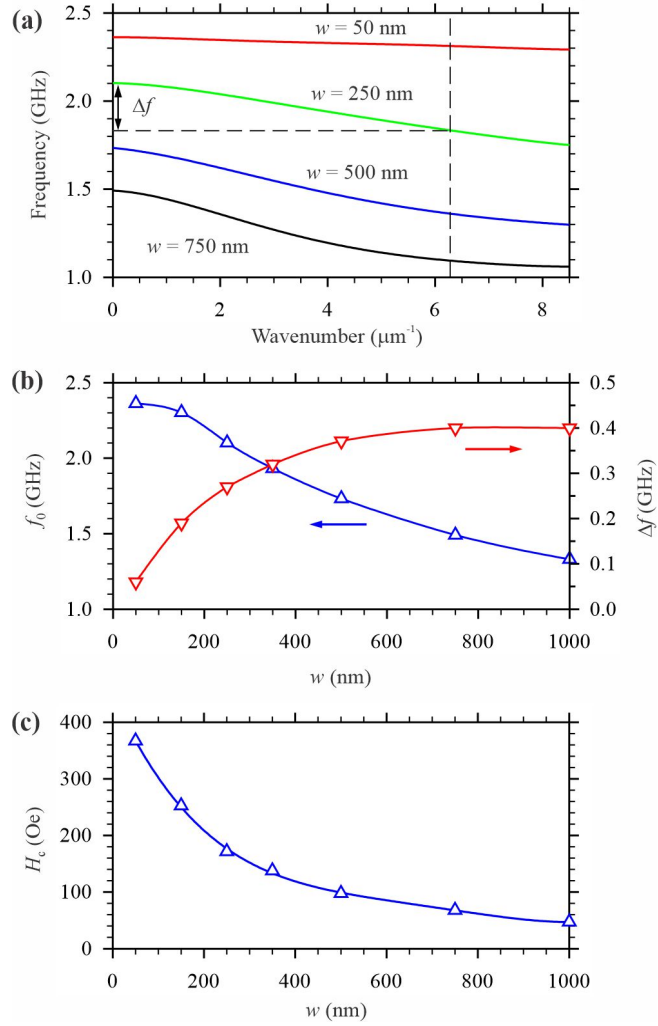

**Figure 4.** Effects of waveguide geometry on propagation of spin waves in zero field. (a) Dispersion curves obtained from micromagnetic simulations at $H = 0$ for waveguides with different widths $w$, as labelled. Vertical dashed line marks the wavelength of 1 μm. $\Delta f$ denotes the spectral width of the spin-wave band, which can be efficiently excited by an inductive antenna. (b) Calculated dependences of the frequency of spin waves at $k = 0$, $f_0$, and the spectral width of the accessible spin-wave band, $\Delta f$, on the width of the waveguide. Curves are guides for the eye. (c) Dependence of the coercive field on the width of the waveguide. Curve is a guide for the eye.